\documentclass[ preprintnumbers,prd,floatfix,nofootinbib,superscriptaddress]{revtex4}
\usepackage{graphicx}

\usepackage[paperwidth=210mm,paperheight=297mm,centering,hmargin=2cm,vmargin=2.5cm]{geometry}

\usepackage[tbtags]{amsmath}  % AMS math 
\usepackage{tabularx}
\usepackage{amssymb}          % AMS symbols
\usepackage{bm}               % bold math
\usepackage{graphicx}         % PostScript figures
\usepackage[export]{adjustbox}% to align images
\usepackage{hhline,multirow}  % for nicer tables
\usepackage{dcolumn}          % Align table columns on decimal point
\usepackage{slashed}
\usepackage{datetime}
\usepackage{color}
\usepackage[dvipsnames]{xcolor}
\usepackage{slashed}
\usepackage{subfloat}
\usepackage{subeqnarray}

\usepackage{amscd}            % for extensible arrows (e.g., limits)
\usepackage{epsfig}
\usepackage{dcolumn}          % Align table columns on decimal point
\usepackage[dvipsnames]{xcolor}
\usepackage[normalem]{ulem}
\usepackage{mathtools,cases}
\usepackage{comment}
\usepackage[utf8]{inputenc}

\RequirePackage{color}
\RequirePackage[colorlinks=true
,urlcolor=black%blue
,anchorcolor=black%blue
,citecolor=black%blue
,filecolor=black%blue
,linkcolor=black%blue
,menucolor=black%blue
,pagecolor=black%blue
,linktocpage=black%true
,pdfproducer=medialab
,pdfa=true
]{hyperref}

\newcommand{\colav}{\frac{\text{Tr}_c}{N_c}}

\newcommand{\Tr}{{\rm Tr}}
\newcommand{\vect}[1]{\boldsymbol{#1}}

\def\beq{\begin{equation}}
\def\eeq{\end{equation}}
\def\bea{\begin{eqnarray}}
\def\eea{\end{eqnarray}}
\def\nn{\nonumber}

\newcommand{\id}{{\mathbb{I}}}
\newcommand{\nocontentsline}[3]{}
\newcommand{\tocless}[2]{\bgroup\let\addcontentsline=\nocontentsline#1{#2}\egroup}

\allowdisplaybreaks[2]

\def\Title#1{\begin{center} {\Large #1 } \end{center}}
\def\Author#1{\begin{center}{ \sc #1} \end{center}}
\def\Address#1{\begin{center}{ \it #1} \end{center}}

\newenvironment{Abstract}{\begin{quotation}  }{\end{quotation}}
\newenvironment{Presented}{\begin{quotation} \begin{center} 
             PRESENTED AT\end{center}\bigskip 
      \begin{center}\begin{large}}{\end{large}\end{center} \end{quotation}}
\begin{document}
\preprint{JLAB-THY-23-3906}
\begin{titlepage}
%\pubblock
\vfill
\Title{Hadronization dynamics from the spectral representation \\ of the gauge invariant quark propagator }
\vfill
%\vspace{-1cm}
\Author{Caroline S. R. Costa}
\Address{Theory Center, Jefferson Lab, 12000 Jefferson Avenue, Newport News, VA 23606, USA}
\Author{Alberto Accardi}
\Address{Hampton University, Hampton, VA 23668, USA \\ 
Theory Center, Jefferson Lab, 12000 Jefferson Avenue, Newport News, VA 23606, USA}
\Author{Andrea Signori}
%\thanks{Electronic address: andrea.signori@unito.it - \href{https://orcid.org/0000-0001-6640-9659}{ORCID: 0000-0001-6640-9659}} 
\Address{Department of Physics, University of Turin, via Pietro Giuria 1, I-10125 Torino, Italy \\
INFN, Section of Turin, via Pietro Giuria 1, I-10125 Torino, Italy}
\vfill
\begin{Abstract}
Using the spectral representation of the quark propagator we study the Dirac decomposition of the gauge invariant quark propagator, whose imaginary part describes the hadronization of a quark as this interacts with the vacuum.

We then demonstrate the formal gauge invariance of the so-called jet mass, that is of the coefficient of the chiral-odd part of the gauge invariant propagator, that can be expressed in any gauge as the first moment of the chiral-odd quark spectral function. This is therefore revealed to be a \textit{bona fide} QCD observable encoding aspects of the dynamical mass generation in the QCD vacuum, and is furthermore experimentally measurable in specific twist-3 longitudinal-transverse asymmetries in DIS and in semi-inclusive electron-positron collisions. In light-like axial gauges, we also obtain a new sum rule for the spectral function associated with the gauge fixing vector.

We finally present a gauge-dependent formula that connects the second moment of the chiral-even coefficient of the quark spectral function to invariant mass generation and final state rescattering in the hadronization of a quark. Finding twist-4 experimental observables sensitive to this quantity is left for future work. 

\end{Abstract}
\vfill
\begin{Presented}
DIS2023: XXX International Workshop on Deep-Inelastic Scattering and
Related Subjects, \\
Michigan State University, USA, 27-31 March 2023 \\
     \includegraphics[width=5.cm]{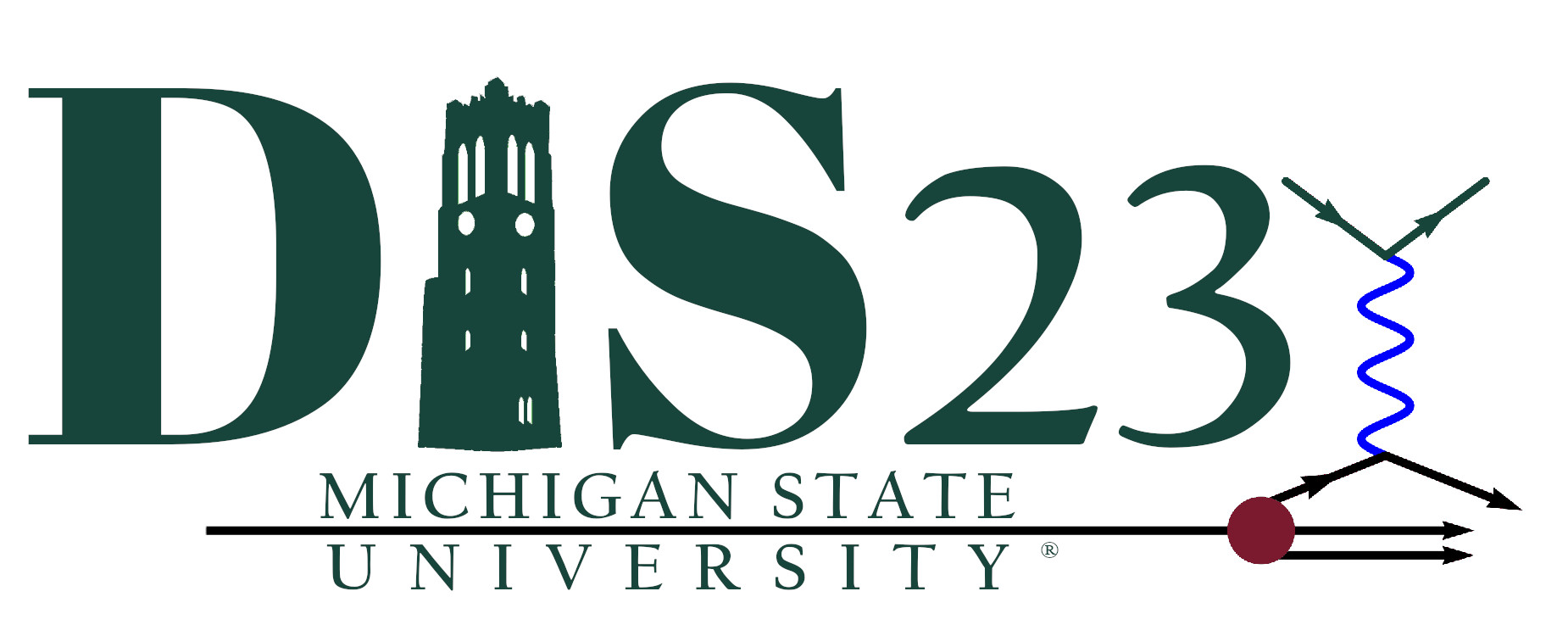}
\end{Presented}
\vfill
\end{titlepage}

%%%%%

\newpage
%%%%%%%%%%%%%%%%%%%%%%%%%%%%%%%%%%%%%%%%%%%%%%%%%%%%%%%%%%%%%%%%%%%%%%%%
\section{Introduction}
\label{s:intro}
%%%%%%%%%%%%%%%%%%%%%%%%%%%%%%%%%%%%%%%%%%%%%%%%%%%%%%%%%%%%%%%%%%%%%%%%%

The dynamics of hadronization is a fundamental aspect in the study of QCD. 
The colored quarks and gluons are the fundamental building blocks of the strong interactions. Yet, even limiting oneself to the light flavor sector, where the current quarks are nearly massless, what one experimentally observes are color neutral and massive hadrons that emerges out of these basic constituents and their interactions.
The exact details of how this process takes place is far from being fully understood, but studying it will shed light on QCD dynamics and
%and hadron formation
its intricate emergent phenomena: color confinement and dynamical mass generation. 

\begin{figure}[b!]
    \centering
    \includegraphics[width=1\textwidth]{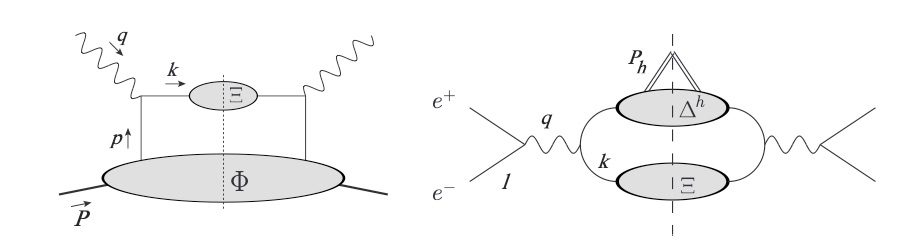}
    \caption{Gauge invariant quark propagator $\Xi$ in inclusive DIS (left) and in $\Lambda$ production from semi-inclusive annihilation (right)~\cite{Accardi:2022oog}.}
    \label{fig:1}
\end{figure}

As proposed in \cite{Accardi:2017pmi, Accardi:2019luo,Accardi:2020iqn,Accardi:2023cmh}, the hadronization process can be studied by analysing the gauge invariant quark propagator, whose imaginary part can be interpreted as the hadronization of a quark as this interacts with the vacuum. 
Not just of mere theoretical interest, the gauge invariant quark propagator and the formalism derived thereof finds phenomenological applications, providing one with observables that can be experimentally accessed in inclusive Deep Inelastic Scattering (DIS) and Semi-Inclusive electron-positron Annihilation (SIA) \cite{Accardi:2017pmi,USBelleIIGroup:2022qro,Accardi:2022oog}.
Indeed, as illustrated in Fig.~\ref{fig:1}, at large values of Bjorken $x$ a gauge invariant quark propagator $\Xi$ replaces the customary free quark line in inclusive DIS and SIA because of kinematical restrictions  \cite{Becher:2006nr,Becher:2006mr,Accardi:2018gmh}. Indeed, at large $x$ the invariant mass of the final state produced by the scattered quark scales as $Q^2(1-x)$, with $Q^2$ the four momentum squared of the exchanged photon, thus limiting the spread of the produced hadrons and collimating these into a jet without the need of selecting jet-like events by a jet finding algorithm as in high energy collisions.

The first diagram in Fig.~\ref{fig:1} has already been used in the context of collinear factorization to determine mass corrections to unpolarized structure funtions and can, likewise, be applied to the polarized $g_1$ structure function \cite{Accardi:2008ne}. 
In Ref.~\cite{Accardi:2017pmi}, DIS with a gauge invariant quark propagator has been analyzed, revealing a novel coupling of the jet mass to the target’s transversity distribution function, which appears in longitudinal-transverse (LT) spin asymmetries and enables one to experimentally study dynamical mass generation effects. An analogous coupling of the jet mass to the chiral-odd component of the self-polarizing $\Lambda$ hadron in electron-positron SIA processes (see the second diagram) has recently been explored in Ref.~\cite{Accardi:2022oog}.

This contribution is a summary of the results discussed in~\cite{Accardi:2023cmh}, and connects them to the relevant experimental observables discussed in Ref.~\cite{Accardi:2017pmi,USBelleIIGroup:2022qro,Accardi:2022oog}. It focuses on the Dirac decomposition of the gauge invariant quark propagator $\Xi$, with particular emphasis on the imaginary part that characterizes the hadronization process, and how this can be expressed in terms of specific moments of the regular quark propagator's spectral functions. 
By leveraging the gauge invariance of $\Xi$, the gauge invariance of the jet mass $M_j \sim \text{Disc} \int dk^+ \Tr({\Xi})$, which is a crucial observable  that quantifies the mass acquired by the quark's fragments during the hadronization process, is demonstrated within a formal framework. Remarkably, the jet mass can be expressed in any gauge as the first moment of the chiral-odd quark spectral function calculated in that gauge. Additionally, a gauge-dependent formula is presented, establishing a connection between the second moment of the chiral-even quark spectral function and the generation of invariant mass as well as the rescattering of final states during the quark's hadronization process. Furthermore, a novel sum rule for the spectral function associated with the gauge fixing vector is derived.
Through these investigations, we aim to shed light on the intricate dynamics underlying hadronization in DIS and its implications for high-energy experiments.

%%%%%%%%%%%%%%%%%%%%%%%%%%%%%%%%%%%%%%%%%%%%%%%%%%%%%%%%%%%%%%%%%%%%%%%%
\section{Gauge invariant quark propagator}
\label{s:axial_gauges}
%%%%%%%%%%%%%%%%%%%%%%%%%%%%%%%%%%%%%%%%%%%%%%%%%%%%%%%%%%%%%%%%%%%%%%%%%

The jet correlator in the particular form that we are interested in this work is given by the discontinuity of the gauge invariant quark propagator written as ~\cite{Accardi:2019luo,Accardi:2020iqn}
\begin{align}
   \Xi_{ij}(k;w) = \mathrm{Disc} \int d^4p  \colav
   \langle\Omega|i\widetilde{S}_{ij}(p;v)\widetilde{W}(k-p;w,v)
   |\Omega\rangle \ .
\label{def:jetcorr}
\end{align}
Here $i\widetilde{S}_{ij}(p,v)$ is a quark bilinear operator whose form will be shortly specified and $\widetilde{W}(k-p;\omega,v)$ is the Fourier transform of a Wilson line that connects the quark fields, providing one with a gauge invariant definition. Furthermore, in Eq.~\eqref{def:jetcorr}, $k$ denotes the quark 4-momentum, $|\Omega\rangle$ is the interacting vacuum state and $w$ is the 4-vector defining the direction of the Wilson line, which in this work is taken to be a staple along the lightcone direction, $w=n_+$. The $4$-vector $v$ appearing in both the quark operator $i\widetilde S$ and in the Wilson line $\widetilde W$ defines an axial gauge, specified by the condition $v\cdot A=0$. 

In the specific case of $v$ being light-like, i.e, $v^2=0$, under Lorentz, parity and time-reversal invariance, the quark operator $i\widetilde S$ has the general form:
\bea
i\widetilde{S}(p, v) = \hat{s}_3(p^2)\slashed{p}+\sqrt{p^2}\hat{s}_1(p^2)\mathbb{I}+\frac{p^2}{p\cdot v} \hat{s}_0(p^2)\slashed{v},
\label{eq:sij_lcg}
\eea
where $\hat{s}_i(p^2)$ are spectral operators that are only functions of $p^2$~\cite{Accardi:2023cmh}. In non-axial gauges, one can simply set $v=0$ in Eq.~\eqref{eq:sij_lcg}. 
The quark operator can also be given an integral representation of the K\"{a}ll\'en-Lehmann type~\cite{Bjorken:1965zz,Weinberg:1995mt} by noticing that 
\bea
    \colav \, \langle\Omega|i\tilde S(p,v)|\Omega\rangle 
    = 
    \int \frac{d\sigma^2}{(2\pi)^4} \frac{\rho_3(p^2) \slashed{p} +\sqrt{p^2}\rho_1(p^2)+ (p^2/p\cdot v)\rho_{0}(p^2)\slashed{v}}{p^2-\sigma^2+i0} \theta(\sigma^2).
\eea
Moreover, one can see that the above expression has a discontinuity $p^2=\sigma^2$ and therefore, each of the spectral functions $\rho_i$ can be obtained by taking the discontinuity of the corresponding spectral operators $\hat{s}_i$:
\bea
    \mathrm{Disc} \colav \, \langle\Omega|\hat{s}_{3,1,0}(p,v)|\Omega\rangle 
    %&=& \
    %frac{1}{(2\pi)^3}\int_{0}^{\infty}d\sigma^2\rho(\sigma^2)\delta(p^2-\sigma^2)\theta(p^-)
    %\nn\\
    &=& \frac{1}{(2\pi)^3}\rho_{3,1,0}(p^2)\theta(p^2)\theta(p^-).
\label{eq:vacdisc}
\eea

The boost along the minus direction effectively provides the inclusive jet correlator with a twist-expansion governed by a power counting scale $\Lambda$:
\begin{align}
    J_{ij}(k^{-},\bm{k_T};n_{+}) 
    &\equiv \frac{1}{2}\int dk^{+}\Xi_{ij}(k;w=n_{+}) \nn\\
    & = \dfrac{1}{2}\alpha(k^-)\gamma^+ + \dfrac{\Lambda}{k^-}\left[\zeta(k^-)\mathbb{I}+\alpha(k^-)
    \dfrac{\vect{\slashed{k}_T}}{\Lambda} \right]+\dfrac{\Lambda^2}{2(k^-)^2}\omega(k^-,\bm{k_T}^2)\gamma^- \ ,
\label{eq:TMDcor}
\end{align}
where $\alpha(k^-)$,  $\zeta(k^-)$ and $\omega(k^-,\bm{k_T}^2)$ are, respectively, the twist-$2$, twist-$3$ and twist-$4$ coefficients of the inclusive jet correlator $J$. These coefficients functions can  be calculated by projecting of $J$ onto a suitable Dirac matrix $\Gamma$ and calculating $J^{[\Gamma]} \equiv \Tr \left[J\, \frac{\Gamma}{2} \right] = \frac{1}{2}\int dk^{+}\Tr{\left[\Xi\, \frac{\Gamma}{2} \right]}$. It is worth mentioning that, because of the integration over $k^+$, in these calculations one only needs to deal with the projections of the Wilson line on the light-cone plus direction and the transverse plane, in which case the Wilson line has the same staple-like shape appearing in TMD factorization theorems, and an explcit formula can be found in Refs.~\cite{Accardi:2020iqn, Accardi:2023cmh}. The result, \textit{in any gauge}, is \cite{Accardi:2023cmh}:
\begin{align}
\alpha(k^-)
&= J^{[\gamma^-]}  = \frac{\theta(k^-)}{2(2\pi)^3} \int_{0}^{\infty} dp^{2} \, \rho_{3} \text(p^2), \\
 \zeta(k^-)
 & =\frac{k^-}{\Lambda}J^{[\mathbb{I}]} = \frac{\theta(k^-)}{2\Lambda(2\pi)^3}\int dp^2 \;\sqrt{p^2}\;\rho_1(p^2),\\
 \omega(k^-,\bm{k_T}) 
 & = \left(\frac{k^-}{\Lambda}\right)^2J^{[\gamma^+]} =
    \frac{\theta(k^-)}{(2\Lambda)^2(2\pi)^3}
    \left( 
    \mu_j^2 +\tau_j^2 + \bm{k_T}^2
    \right) \ .
    \label{eq:omega}
\end{align}
In the second line the jet mass is formally expressed as
\begin{align}
    M_j = \int_{0}^{\infty} dp^2 \;\sqrt{p^2}\;\rho_1(p^2) \ ,
\end{align}
a form that justifies its physical interpretation as the average mass of the particles produced in the hadronization process. Note that the $\rho_1$ function is itself gauge dependent, but not so its first moment.
In the last line, 
\begin{equation}
    K_j^2 = \mu_j^2 +\tau_j^2
    \label{eq:Kj2}
\end{equation}
is the jet virtuality. Were the quark on its mass shell, that is if one considered a quark line instead of the jet correlator $\Xi$, one would have $K_j^2=m^2$. In our case, with a final state jet produced during the hadronization process, $K_j^2 > m_q^2$ is of hadronic size, and receives contributions both from the ``bare'' invariant mass $\mu^2_j$ initially produced during the quark fragmentation process, and from the transverse broadening $\tau_j^2$ acquired by the fragmentation products due to final state interactions. The bare $\mu_j^2$ can be written in a simple form in terms of the second moment of the chiral-even quark spectral function,
\begin{align} 
    \mu_j^2 = \int_0^\infty dp^2 \,p^2\, \rho_3(p^2) ,
\end{align}
which makes explicit its physical interpretation even though (unlike $M_j$) its numerical value is not gauge invariant. Only the sum of $\mu_j^2$ and $\tau_j^2$ is gauge invariant. However, the latter does not find a simple form in terms of the quark spectral functions, and we leave it to the next section to expose this explicitly. Using the results obtained for the $\alpha$, $\zeta$ and $\omega$ coefficients, the jet correlator in~\eqref{eq:TMDcor} has the final form
\begin{align}
\label{e:J_Dirac_explicit}
  J(k^-,\bm{k_T};n_+)
    = \frac{\theta(k^-)}{4(2\pi)^3\, k^-} \, 
    \bigg\{ k^-\, \gamma^+ + \slashed{k}_T + M_j \id + \frac{K_j^2 + \bm{k_T}^2}{2k^-} \gamma^- \bigg\} \ .
\end{align}
One recognizes the similarity to the free quark propagator
%\begin{equation}
$
    \slashed{k} + m = k^-\, \gamma^+ + \slashed{k}_T + m \id
    +  \frac{m^2 + \bm{k_T}^2}{2k^-}\, \gamma^- \, ,
$
%\end{equation}
with the jet mass $M_j$ generalizing the bare quark mass $m$, and the jet virtuality $K_j^2$ replacing the on-mass-shell $m_q^2$ as discussed.

%%%%%%%%%%%%%%%%%%%%%%%%%%%%%%%%%%%%%%%%%%%%%%%%%%%%%%%%%%%%%%%%%%%%%%%%
\section{Quark spectral functions sum rules}
\label{s:2}
%%%%%%%%%%%%%%%%%%%%%%%%%%%%%%%%%%%%%%%%%%%%%%%%%%%%%%%%%%%%%%%%%%%%%%%%%

The jet correlator $J$ is gauge invariant, and so are its $\alpha$, $\zeta$, and $\omega$ coefficients. For the same reason, $J$ is independent of the axial-gauge-fixing vector $v$ should one choose a light-like gauge. Calculations of the jet function projections were then carried out in \cite{Accardi:2023cmh} both in light-cone and generic gauges. As a result, one obtains that the following sum rules for the quark spectral functions are satisfied independently of the chosen gauge: 
    \begin{align}
    1 & =\int_{0}^{\infty}dp^{2}\rho_3(p^2),
    \label{eq:rho3sumrule}
    %& = \frac{\theta(k^-)}{2(2\pi)^3},
    \\
    M_j &= \int_{0}^{\infty} dp^2 \;\sqrt{p^2}\;\rho_1(p^2),
     \label{eq:rho1sumrule}\\
    0 &=\int_{0}^\infty dp^2\,p^2\, \rho_0(p^2).
    \label{eq:rho0sumrule}
    \end{align}
It is worth emphasizing that although the quark spectral functions are themselves gauge dependent, the sum rules above are independent of the chosen gauge. Eq.~\eqref{eq:rho3sumrule}, in particular, is already known to be independent of the chosen gauge, since it follows from the canonical commutation relations \cite{Weinberg:1995mt}. It has been formally verified in~\cite{Accardi:2023cmh} resorting to techniques that also allowed to obtain the jet mass sum rule \eqref{eq:rho1sumrule} and the axial spectral function sum rule \eqref{eq:rho0sumrule} which are new results.

All these sum rules can be an important tool to verify actual calculations of the quark propagator, such as in Ref.~\cite{Horak:2022aza,Falcao:2022gxt,Solis:2019fzm,Duarte:2022yur} and references therein\footnote{Those calculations, however, have only been carried out in a covariant gauge. Ideally, one would also want a light cone gauge calculation to explicitly verify the gauge invariance of the sum rules, as well as investigate the novel axial gauge sum rule (\ref{eq:rho0sumrule}).}, but more importantly the jet mass sum rule can be considered the most remarkable and interesting one because (i) it provides one with a gauge invariant and nonperturbative mass that is dynamically generated and that can be interpreted as the mass acquired by the quark during its hadronization \cite{Accardi:2019luo}, and (ii) it can be experimentally accessed in inclusive DIS and semi-inclusive $e^{+}e^{-}$ annihilation~\cite{bauer2023numerical}. Indeed, in inclusive DIS, $M_j$ couples to the chiral-odd transversity distribution $h_1(x)$ of the proton target and contributes to the twist-3 LT double spin asymmetry~\cite{Accardi:2017pmi}. 
In electron-positron SIA with an observed self-polarizing $\Lambda$ hadron, the jet mass $M_j$ similarly contributes to the LT double spin asymmetry but coupled with the chiral-odd twist-3 $H_1^\Lambda(z)$ fragmentation function of $\Lambda$ \cite{USBelleIIGroup:2022qro,Accardi:2022oog}.

The formalism deployed in the derivation of the spectral sum rules \eqref{eq:rho0sumrule}-\eqref{eq:rho3sumrule} can also be used for the calculation of the gauge invariant twist-4 coefficient $\omega$ and the corresponding jet virtuality $K_j^2=\mu_j^2+\tau_j^2$ discussed in the previous section. While $K_j^2$ is gauge invariant, its decomposition in invariant mass $\mu_j^2$ of the hadronization products and transverse momentum broadening $\tau_j^2$ in the final state given in Eq.~(\ref{eq:Kj2}) is not. To see that in detail, it is convenient now to look at the explicit form of $\tau_j^2$ which had been postponed in the previous section:
\begin{align}
    \tau_j^2 
        &= (2\pi)^3 \int_{0}^{\infty} dp^2 \,\mathrm{Disc}\frac{\mathrm{Tr_{c}}}{\mathrm{N_c}}\langle\Omega|\hat{\sigma}_3(p^2)ig \,\vect{D_T} \big[ \vect{A_T}(\bm{\xi_T})+\vect{\mathcal{Z}_T}(\bm{\xi_T}) \big]_{\bm{\xi_T}=0}|\Omega\rangle,
\label{eq:tauj_general_gauge}
\end{align}
with
\bea
    \vect{\mathcal{Z}_T} = ig \!
    \int_{0}^{\infty^+} \hspace*{-0.5cm} ds^+ \mathcal{U}_{n_{+}}[0^-,0^+,\vect{\xi_{T}};0^-,s^+,\vect{\xi_{T}}]\bm{G}^{\bm T -} (0^-,s^+,\vect{\xi_{T}}) \, 
    \mathcal{U}_{n^+}[0^-,s^+,\vect{\xi_{T}};0^-,\infty^+,\vect{\xi_{T}}]|\Omega\rangle.
\label{eq:final-state-interactions}
\eea
Here, $\vect{D_{T}}= \vect{\partial_T}+ ig \vect{A_{T}}$ is the transverse covariant derivative, $\bm{G^{T -}}$ is the field strength tensor, with repeated transverse T indexes contracted in the 2D transverse Euclidean space, and $\mathcal{U}_v [a;\infty]$ is a straight Wilson line from $a$ to infinity along the $v$ direction,
%
%\begin{align}
$    \mathcal{U}_v [a;\infty] 
    = 
    \mathcal{P}\exp{\left(-ig\int_{0}^{\infty}ds\, v^{\mu}A_{\mu}\big(a+sv) \big)\right)}.
$
%\end{align}
%
The $\vect{D_{T}}$ acting on the square bracket in Eq.~\eqref{eq:tauj_general_gauge} introduces one further insertion of $\bm{G}^{\bm T -}$ inside the $\cal U$ Wilson line providing one with the basis for the interpretation of $\tau_j^2$ as a final state rescattering term.  

In a generic gauge $\tau_j^2 \neq 0$. In the the light-cone gauge, however, 
one notices that $\bm{G}^{\bm T -}=-\partial^{-} {\bf A}_{T}$, and imposing advanced boundary conditions (i.e., $\vect{A}_{T}(\infty)=0)$ one obtains a vanishing $(\tau_j^{2})^\mathrm{lcg}=0$: no rescattering seems to occur in the light cone gauge. 
%Care should be taken, therefore, before concluding that no final state interactions are present in the light-cone gauge. 
Although final state interactions seemingly disappear with the appropriate choice of gauge and boundary conditions, their effect is actually embedded inside $(\mu_j^2)^\text{lcg}$, more specifically inside the chiral-even quark spectral function $\rho_3^\text{lcg}$ which is not gauge invariant and absorbs the rescattering effects explicitly occurring in other gauges.

%%%%%%%%%%%%%%%%%%%%%%%%%%%%%%%%%%%%%%%%%%%%%%%%%%%%%%%%%%%%%%%%%%%%%%%%%

%%%%%%%%%%%%%%%%%%%%%%%%%%%%%%%%%%%%%%%%%%%%%%%%%%%%%%%%%%%%%%%%%%%%%%%%%
%%%%%%%%%%%%%%%%%%%%%%%%%%%%%%%%%%%%
\section{Summary}
\label{s:intro}
%%%%%%%%%%%%%%%%%%%%%%%%%%%%%%%%%%%%%%%%%%%%%%%%%%%%%%%%%%%%%%%%%%%%%%%%%

In this proceeding, we have discussed the spectral decomposition \eqref{def:jetcorr}-\eqref{eq:sij_lcg} of the gauge-invariant quark propagator $\Xi$. This was then utilized to calculate the Dirac coefficients of the gauge-invariant inclusive jet correlator $J \sim \int dk^+ \Xi$ in terms of moments of the spectral functions of the quark's Feynman propagator, which is of phenomenological interest since it enters the calculation of DIS and SIA cross sections at large values of the Bjorken $x$ variable, see Fig.~\ref{fig:1}. The gauge invariance of the $\alpha(k^-)$ and $\zeta(k^-)$ twist-2 and twist-3 coefficients of $J$ then leads to \textit{novel sum rules for the quark spectral functions}, see Eqs.~\eqref{eq:rho0sumrule}-\eqref{eq:rho3sumrule}. 

In particular, the jet mass $M_j$, which is an {\em experimentally observable} dynamically generated quark mass, is equal to the first moment of the chiral-odd spectral function $\rho_1$ {\em in any gauge}, even though the spectral function itself depends on the chosen gauge. 
Furthermore, the second moment of the additional quark spectral function $\rho_0$ present in axial gauges must also vanish identically {\em in any gauge}. We look forward to independent investigations and numerical confirmations of these two new sum rules.

The twist-4 coefficient $\zeta$ has also been expressed, but in a gauge dependent way, in terms of the invariant mass of the hadronizing quark and the effect of final-state interactions. This is also, in principle, an experimentally observable quantity that contributes to DIS and SIA processes at twist-4. Looking for specific twist-4 measurements that elucidate the dynamics of invariant mass generation in the QCD fragmentation of a quark is left for future work.

%%%%%%%%%%%%%%%%%%%%%%%%%%%%%%%%%%%%%%%%%%%%%%%%%%%%%%%%%%%%%%%%%%%%%%%%%
%\newpage
\begin{acknowledgments}
This work was supported in part by the  U.S. Department of Energy (DOE) contract DE-AC05-06OR23177, under which Jefferson Science Associates LLC manages and operates Jefferson Lab. 
AA also acknowledges support from DOE contract DE-SC0008791.
AS received support from the European Commission through the Marie Sk\l{}odowska-Curie Action SQuHadron (grant agreement ID: 795475). 
\end{acknowledgments}

%%%%%%%%%%%%%%%%%%%%%%%%%%%%%%%%%%%%%%%%%%%%%%%%%%%%%%%%%%%%%%%%%%%%%%%%%
%%		BIBLIOGRAPHY
%%%%%%%%%%%%%%%%%%%%%%%%%%%%%%%%%%%%%%%%%%%%%%%%%%%%%%%%%%%%%%%%%%%%%%%%%
%\newpage
%%	choose a bibtex style 
\bibliographystyle{JHEP}  % ama, nar, alpha, plain, chicago, abbrv, siam, JHEP
\bibliography{main.bib}
%%%%%%%%%%%%%%%%%%%%%%%%%%%%%%%%%%%%%%%%%%%%%%%%%%%%%%%%%%%%%%%%%%%%%%%%%
 
\end{document}